\documentclass[12pt,a4paper]{article}
\date{\today}

\newcommand{\Tr}{{\rm Tr}}
\newcommand{\insertplot}[5]{\begin{figure}
 \hfill\hbox to 0.05in{\vbox to #5in{\vfill
 \inputplot{#1}{#4}{#5}}\hfill}
 \hfill\vspace{-.1in}
 \caption{#2}\label{#3}
 \end{figure}}
 \newcommand{\inputplot}[3]{
 \special{ps: plotfile #1}
}
\newcounter{fig}   \newcommand{\lbfig}[1]{\refstepcounter{fig}
\label{#1} }
\usepackage{subfigure}
\usepackage{epsfig}
\usepackage{amsmath}
\usepackage{amsfonts}
\usepackage{graphicx}
\usepackage[german, english]{babel}
\usepackage{a4wide}
\usepackage{amsmath}
\usepackage{amssymb}
\usepackage{ifthen}
\usepackage{epsfig}

\begin{document}

\newcommand{\dd}{\mbox{d}}
\newcommand{\tr}{\mbox{tr}}
\newcommand{\la}{\lambda}
\newcommand{\ta}{\theta}
\newcommand{\f}{\phi}
\newcommand{\vf}{\varphi}
\newcommand{\ka}{\kappa}
\newcommand{\al}{\alpha}
\newcommand{\ga}{\gamma}
\newcommand{\de}{\delta}
\newcommand{\si}{\sigma}
\newcommand{\bomega}{\mbox{\boldmath $\omega$}}
\newcommand{\bsi}{\mbox{\boldmath $\sigma$}}
\newcommand{\bchi}{\mbox{\boldmath $\chi$}}
\newcommand{\bal}{\mbox{\boldmath $\alpha$}}
\newcommand{\bpsi}{\mbox{\boldmath $\psi$}}
\newcommand{\brho}{\mbox{\boldmath $\varrho$}}
\newcommand{\beps}{\mbox{\boldmath $\varepsilon$}}
\newcommand{\bxi}{\mbox{\boldmath $\xi$}}
\newcommand{\bbeta}{\mbox{\boldmath $\beta$}}
\newcommand{\ee}{\end{equation}}
\newcommand{\eea}{\end{eqnarray}}
\newcommand{\be}{\begin{equation}}
\newcommand{\bea}{\begin{eqnarray}}
\newcommand{\ii}{\mbox{i}}
\newcommand{\e}{\mbox{e}}
\newcommand{\pa}{\partial}
\newcommand{\Om}{\Omega}
\newcommand{\vep}{\varepsilon}
\newcommand{\vphi}{\varphi}
\newcommand{\bfph}{{\bf \phi}}
\newcommand{\lm}{\lambda}
\def\theequation{\arabic{equation}}
\renewcommand{\thefootnote}{\fnsymbol{footnote}}
\newcommand{\re}[1]{(\ref{#1})}
\newcommand{\R}{{\rm I \hspace{-0.52ex} R}}
\newcommand{\N}{{\sf N\hspace*{-1.0ex}\rule{0.15ex}%
{1.3ex}\hspace*{1.0ex}}}
\newcommand{\Q}{{\sf Q\hspace*{-1.1ex}\rule{0.15ex}%
{1.5ex}\hspace*{1.1ex}}}
\newcommand{\C}{{\sf C\hspace*{-0.9ex}\rule{0.15ex}%
{1.3ex}\hspace*{0.9ex}}}
\newcommand{\eins}{1\hspace{-0.56ex}{\rm I}}
\renewcommand{\thefootnote}{\arabic{footnote}}

\title
{\vskip 2cm Sphaleron solutions of the Skyrme model from Yang-Mills holonomy}
\author{
Ya Shnir and G Zhilin\\[10pt]
{\small Department of Theoretical Physics and Astrophysics}\\
{\small Belarusian State University, Minsk}}
\date{~}
\maketitle

\begin{abstract}
{\sf We discuss how an approximation to the axially symmetric sphalerons in the Skyrme
model can be constructed from the holonomy of a non-BPS Yang-Mills calorons.
These configurations, both in the Skyrme model and in the Euclidean  Yang-Mills theory, are
characterized by two integers $n$ and $m$, where $\pm n$ are the
winding numbers of the constituents and the second integer
$m$ defines type of the solution, it has zero topological charge for even $m$ and for odd values of $m$ the corresponding
chain has topological charge $n$.
It is found numerically that the holonomy of the
chains of interpolating calorons--anticalorons  provides a reasonably good approximation
to the corresponding Skyrmion--antiSkyrmion chains when the topological charge of the
Skyrmion constitutents is two times more than the Chern-Pontryagin index of the caloron. }
\end{abstract}


\section{Introduction}
There is a very interesting interplay between different topological solitons in
 $d=3+1$ and $d=4$ dimensions. It was pointed out by M.~Atiyah and N.~Manton
that the holonomy of Yang-Mills instantons in $d=4$ provides a
very good approximations to Skyrmion solutions of the Skyrme model
in $d=3+1$ \cite{Atiyah:1989}, the energy of the lower charge
$B=1$ Skyrmion is reproduced within a percent accuracy. The
remarkable exact topological results is that if the instanton
holonomy is considered as a Skyrme field, the baryon number $B$
becomes exactly equal to the  Chern-Pontryagin charge $N$.
Furthermore, a suitable parametrization of the instanton solution
yields also the symmetry of the Skyrme field of the higher charges
\cite{Leese:1994,Singer:1999,Sutcliffe:2004,Manton:1995}. This
analogy is especially noticeable because, unlike instantons, the
Skyrmion solutions are not Bogomolnyi-Prasad-Sommerfeld (BPS) states, the energy of the  $B=1$
Skyrmion exceeds the topological energy bound by about 23$\%$.


This method can be extended to set a similar link between Skyrmions and calorons
which are periodic instantons at finite temperature
\cite{Nowak:1989gw,Dey:1994by}. Recently the caloron generated field on
$S^1\times {\mathbb{R}}^3$ was used to
construct axially symmetric Skyrmion chains \cite{Harland:2008}, in this approach
the periodicity of a chain
of Skyrmions matches the period of the gauge field.

Interestingly, if
the time-periodic array of instantons corresponds to the non-trivial holonomy,
the caloron configurations may possess constituent structure by itself  \cite{Baal}.
As the size of the caloron is getting
larger than the period $T$, the caloron is splitting into
constituents, which represent monopole-antimonopole pair
configuration.
On the other hand, the BPS monopole solution of the Yang-Mills-Higgs theory in  $d=3+1$
is equivalent to an infinite chain of instantons directed along the Euclidean time axis
\cite{Rossi:1979}. Thus, it is possible to
construct a family of calorons which interpolates between the monopoles at one end and the instantons
at the other end \cite{Ward:2003sx}.

Another similarity between the Skyrme model, Yang-Mills theory in $d=4$ and
Yang-Mills-Higgs theory in $d=3+1$ is that all these models support axially symmetric sphaleron
solutions which represent chains of interpolating
Skyrmion--antiSkyrmions \cite{Krusch:2004uf,Shnir:2009ct},
non-self-dual instantons \cite{Radu:2006gg,Shnir:2007zz} and non-BPS monopole--antimonopole chains
consisting of a charge $n$ and a charge $-n$ constituents \cite{KKS},
respectively.
Since these sphaleron configurations, both in the Yang-Mills model, Yang-Mills-Higgs model
and in the Skyrme model, are deformations of topological sectors
with a given charge, the statement about the topological
equivalence of the instanton induced holonomy and the
corresponding Skyrme field cannot be applied straightforwardly.

The purpose of this letter is to compute the holonomy of the
axially-symmetric caloron-anti-caloron configurations and compare
it to the Skyrmion--antiSkyrmion chains. We show that it provides
rather good approximation to the sphaleron solutions of the Skyrme
model if the baryon number of the constituents in the latter
sector is taken to be twice as much as the topological charge of
the Yang-Mills calorons. This result is not very surprising
because it is known that the the Skyrmion--antiSkyrmion chains may
exist only for values $B\ge 2$ \cite{Krusch:2004uf,Shnir:2009ct}.
It was also pointed out that the transition between the chains and
ring-like configurations for the Skyrmion--antiSkyrmion pair it
taken place as the charge $B$ increases above $B=4$
\cite{Krusch:2004uf}, in the case of monopole-antimonopole pair
similar transition occurs above $N=2$ \cite{KKS}, this indicates
that the gauge interaction between the constituents in the
monopole-antimonopole pair is much stronger than the dipole-dipole
interaction in the Skyrme--antiSkyrme pair
\cite{Krusch:2004uf,Shnir:2009ct}.

In the next section we discuss the axially symmetric ansatz which we apply to parameterize
the action of the Euclidean Yang-Mills theory, and
the boundary conditions imposed to get regular solution.
The numerical results are presented in
Section 3 where we evaluated the holonomy of the caloron--anticaloron chains and compare it to the numerical
solutions of the Skyrme model.
We give our conclusions and remarks in the final section.

\section{Sphaleron calorons in Euclidean $SU(2)$ Yang-Mills}
As a starting point we consider the  $SU(2)$ Yang-Mills action in $R^3 \times S^1$ with one periodic dimension $x_0 \in [0,T]$
\begin{equation} \label{S}
S = \frac{1}{2}  \int d^4 x \Tr\left( F_{\mu\nu} F_{\mu\nu}\right)
= \frac{1}{4}  \int d^4 x \left(  F_{\mu\nu} \pm {\widetilde  F}_{\mu\nu}\right)^2
\mp \frac{1}{2}  \int d^4 x \Tr\left(  F_{\mu\nu}{\widetilde  F}_{\mu\nu}\right)
\end{equation}
Here $
F_{\mu\nu} = \partial_\mu A_\nu - \partial_\nu A_\mu + i[A_\mu, A_\nu]$ are the components of the $su(2)$-valued field strength
and the topological charge is defined as
\be \label{N}
N= \frac{1}{32\pi^2} \varepsilon_{\mu\nu\rho\sigma} \int d^4x
\Tr  F_{\mu\nu} F_{\rho\sigma}
\ee
A lower bound on the action then is  $S \ge 8 \pi^2 N$, it is saturated by self-dual configurations.

As discussed in \cite{Shnir:2007zz}, non-self dual axially symmetric regular caloron solutions in this theory
can be constructed using the Ansatz for the gauge field
\begin{eqnarray} \label{ansatz}
A_k dx^k
 =
\left( \frac{K_1}{r} dr + (1-K_2) d\theta\right) \frac{\tau_\varphi^{(n)}}{2e}
&-&n \sin\theta \left(( K_3 \frac{\tau_r^{(n,m)}}{2e}
                     +(1-K_4) \frac{\tau_\theta^{(n,m)}}{2e}\right) d\varphi;
\nonumber \\
A_0 = A_0^a\frac{\tau^a}{2}
& = &
\left(K_5 \frac{\tau_r^{(n,m)}}{2}+ K_6\frac{\tau_\theta^{(n,m)}}{2} \right) \  ,
\nonumber
\end{eqnarray}
which was previously applied in the Yang-Mills-Hiigs theory
to obtain various monopole-antimonopole chains and vortex-like configurations \cite{KKS}.
The Ansatz is written in the basis of $su(2)$ matrices
$\tau_r^{(n,m)},\tau_\theta^{(n,m)} $ and
$\tau_\vphi^{(n)}$ which are defined as the dot product of the Cartesian vector of Pauli
matrices $\vec \tau $ and the spacial unit vectors
\begin{eqnarray}
{\hat e}_r^{(n,m)} & = & \left(
\sin(m\theta) \cos(n\vphi), \sin(m\theta)\sin(n\vphi), \cos(m\theta)
\right)\ , \nonumber \\
{\hat e}_\theta^{(n,m)} & = & \left(
\cos(m\theta) \cos(n\vphi), \cos(m\theta)\sin(n\vphi), -\sin(m\theta)
\right)\ , \nonumber \\
{\hat e}_\vphi^{(n)} & = & \left( -\sin(n\vphi), \cos(n\vphi), 0 \right)\ ,
\label{unit_e}
\end{eqnarray}
respectively. The functions $K_i$, $i=1,\dots,6$
depend on the coordinates $r$ and $\theta$. The solutions are labeled by two integers $n,m$ and the
topological charge of configurations is $
N = \frac{n}{2} \left[1-(-1)^m\right]$. Evidently, for even $m$ the configuration has zero topological charge
while for odd $m$ is has a charge $n$. Thus, the solutions represent periodic
arrays of the calorons and the anticalorons of topological charge $\pm n$,
which are located on the axis of symmetry in alternating order \cite{Shnir:2007zz}. For configuration with
$m\ge 2$ the energy bound cannot be attained.

The finiteness of the Euclidean action \eqref{S} requires
$\Tr (F_{\mu\nu} F_{\mu\nu}) \to O(r^{-4})$ as $r \to \infty$.
In the regular gauge the value of the component of the gauge
potential $A_0$  approaches a constant as $r\to \infty$, i.e.,
\be
A_0 \to \frac{i \beta}{2}\tau_r^{(n,m)}
\ee
where $\beta \in [0:2\pi/T]$ and $T$ is the period in the imaginary time direction.
Using the classical scale invariance we can fix $\beta =1$.

\begin{figure}
\lbfig{f-4}
\begin{center}
\includegraphics[height=.30\textheight, angle =0]{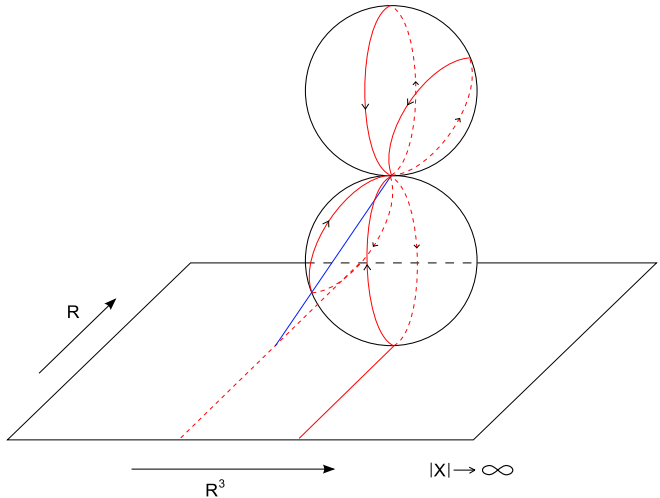}
\end{center}
\caption{Construction of the Skyrmion--antiSkyrmion pair from holonomy.}
\end{figure}

Note that these configurations should be more akin to the monopole--antimonopole chains \cite{KKS},
rather than the instanton–-anti--instantons sphalerons discussed in \cite{Radu:2006gg}, indeed
there is a caloron–-anticaloron chain consisting of a charge $n=\pm 1$ constituents whereas
the corresponding instanton–-anti-instanton chain may exist only for values $n \ge 2$.
These configurations remain above the self-duality bound.

\section{Skyrmion-antiSkyrmion chains from calorons}

We discuss now, how an approximation to the sphalerons in the Skyrme model
can be obtained from the holonomy of a suitable non-self dual calorons.
The Atiyah-Manton approach of generating approximate Skyrmion configurations
on ${\mathbb R}^3$ is to evaluate the holonomy of the corresponding Yang-Mills instantons
along the lines parallel to the Euclidean time axis.
The construction is topologically natural, since the holonomy of an instanton with topological charge $N$
is a Skyrmion with baryon charge $B=N$ \cite{Atiyah:1989}.

\begin{figure}
\lbfig{f-2}
\begin{center}
\subfigure[n=1]{\includegraphics[scale=0.25,angle=-90]{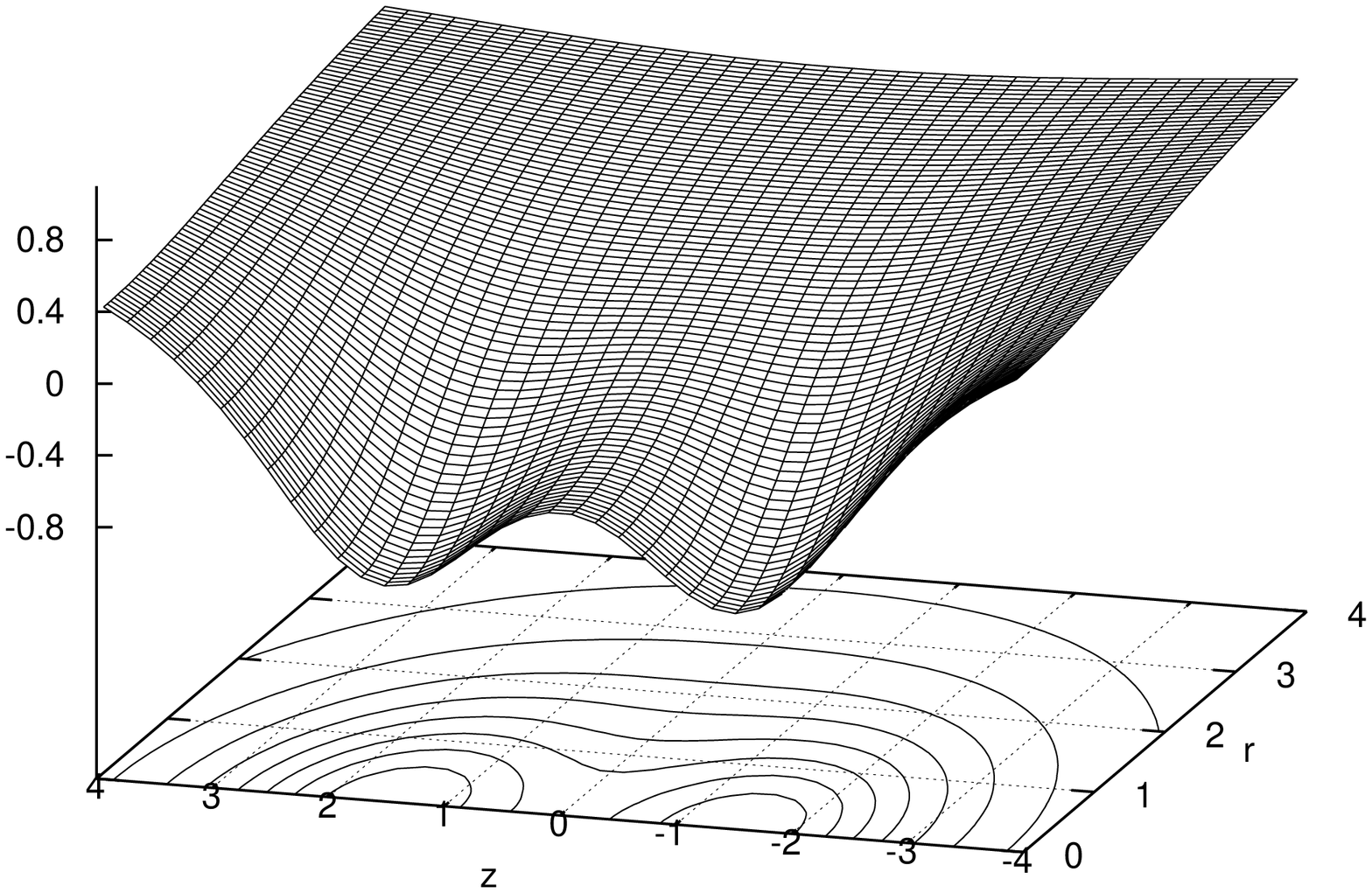}}
\subfigure[$B=2n=2$]{\includegraphics[scale=0.25,angle=-90]{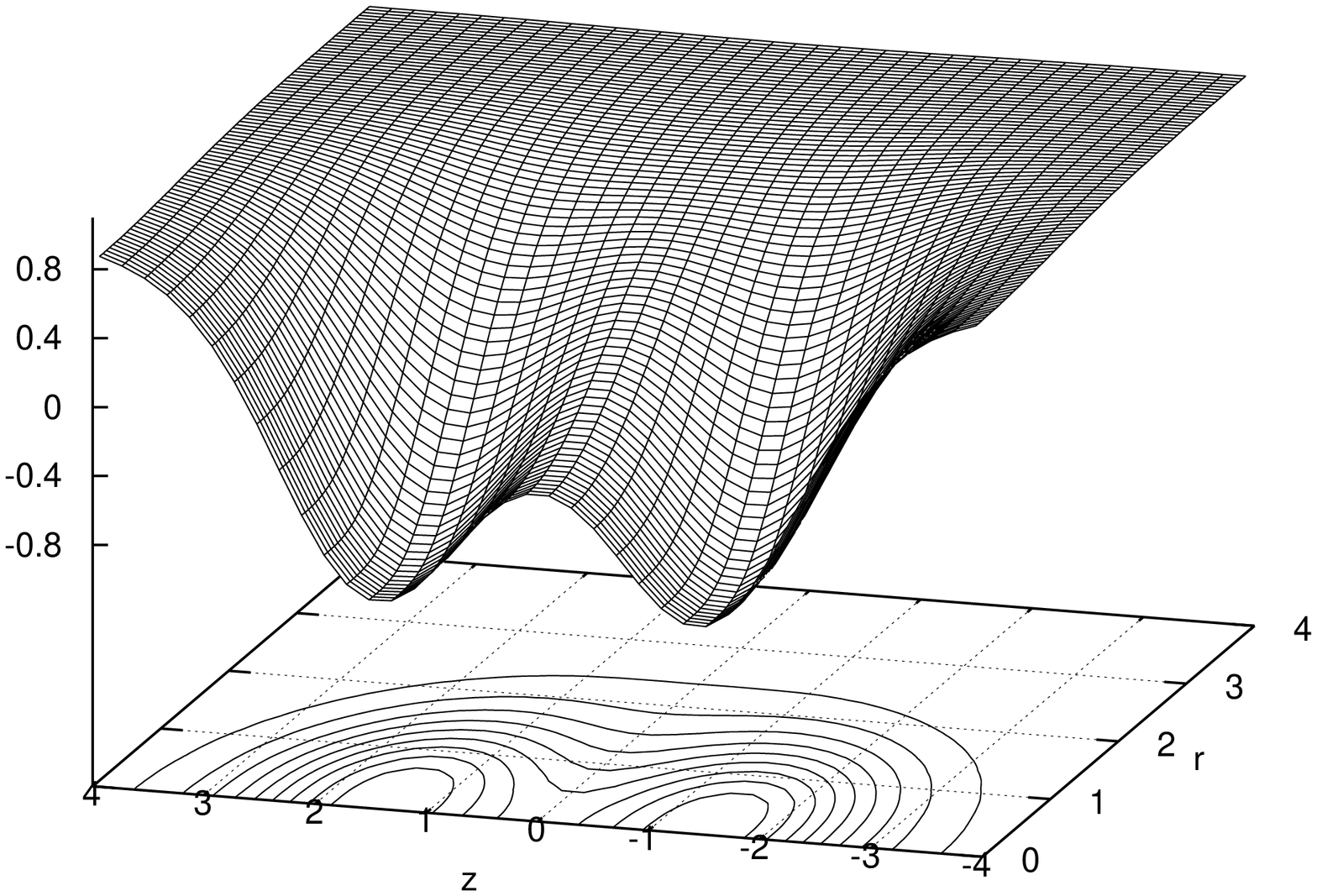}}
\subfigure[n=2]{\includegraphics[scale=0.25,angle=-90]{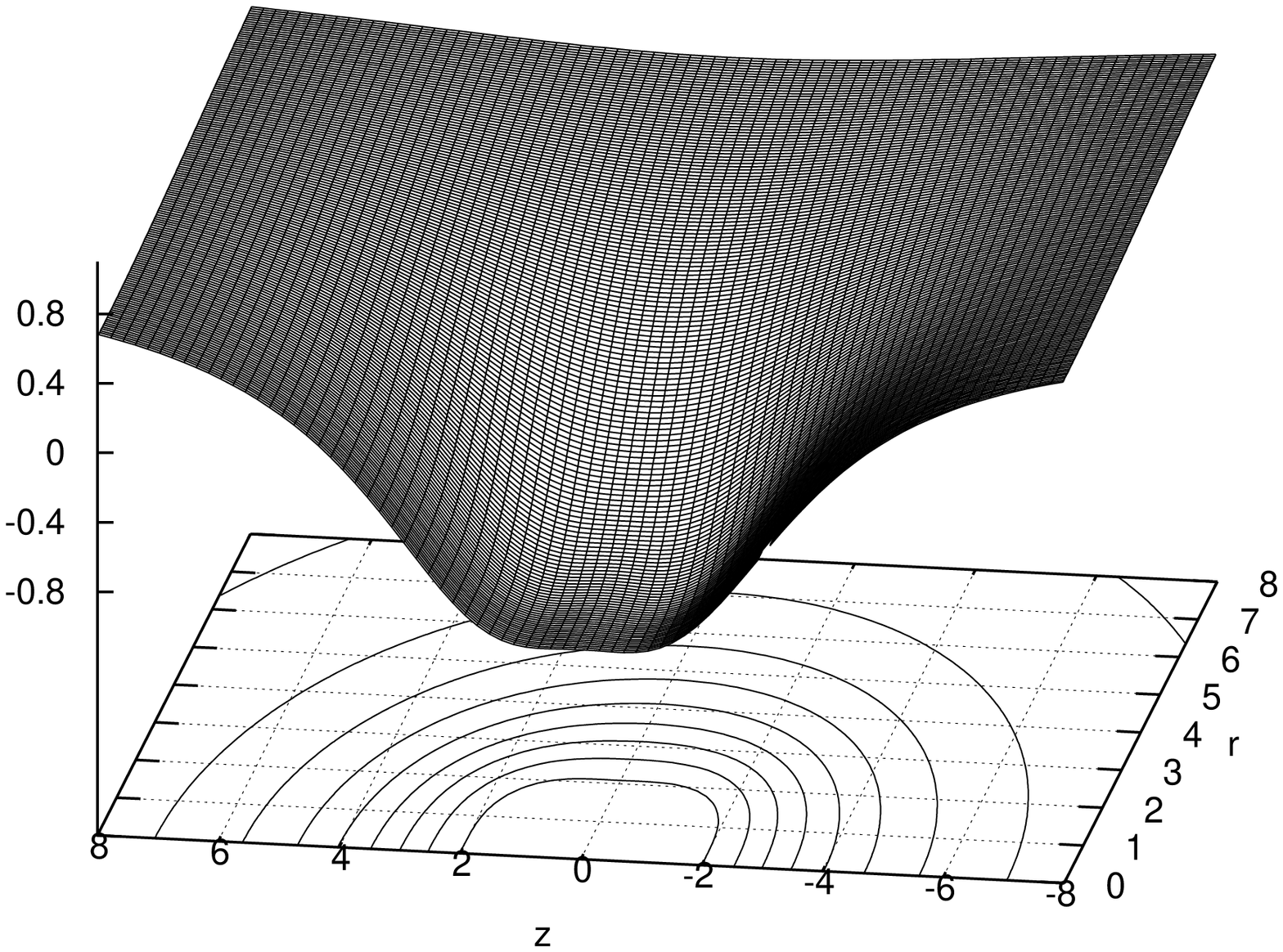}}
\subfigure[$B=2n=4$]{\includegraphics[scale=0.25,angle=-90]{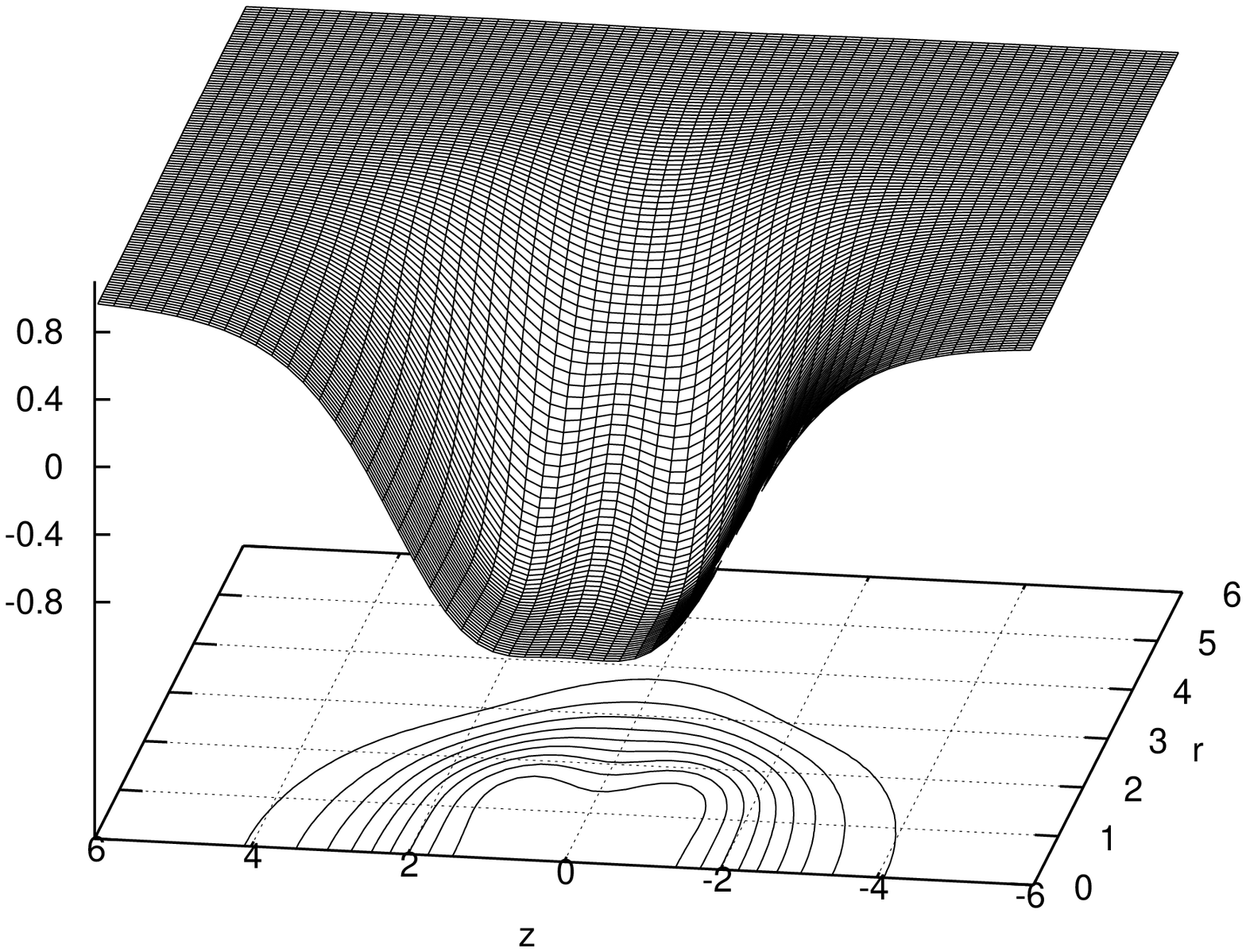}}
\subfigure[n=3]{\includegraphics[scale=0.25,angle=-90]{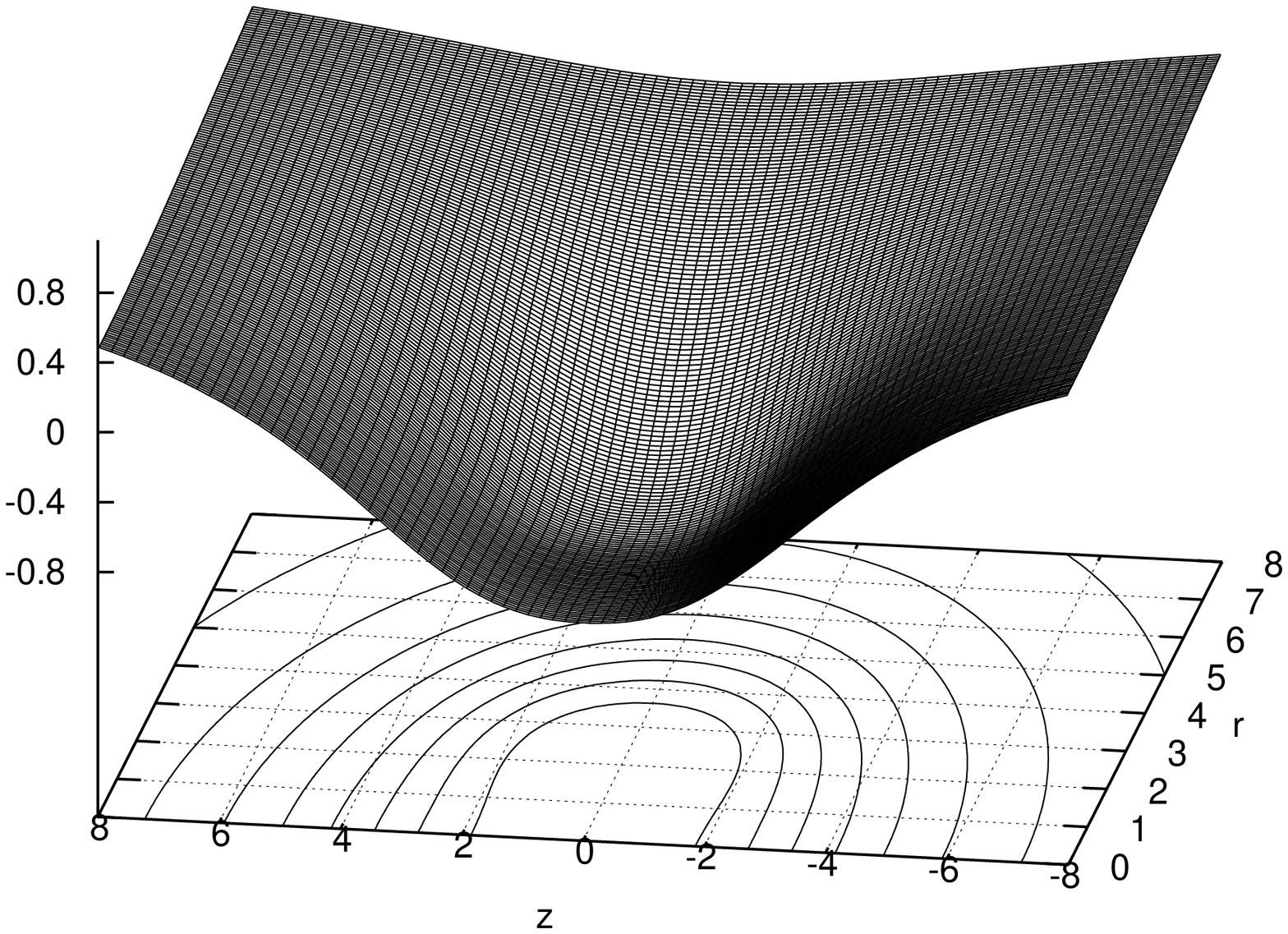}}
\subfigure[$B=2n=6$]{\includegraphics[scale=0.25,angle=-90]{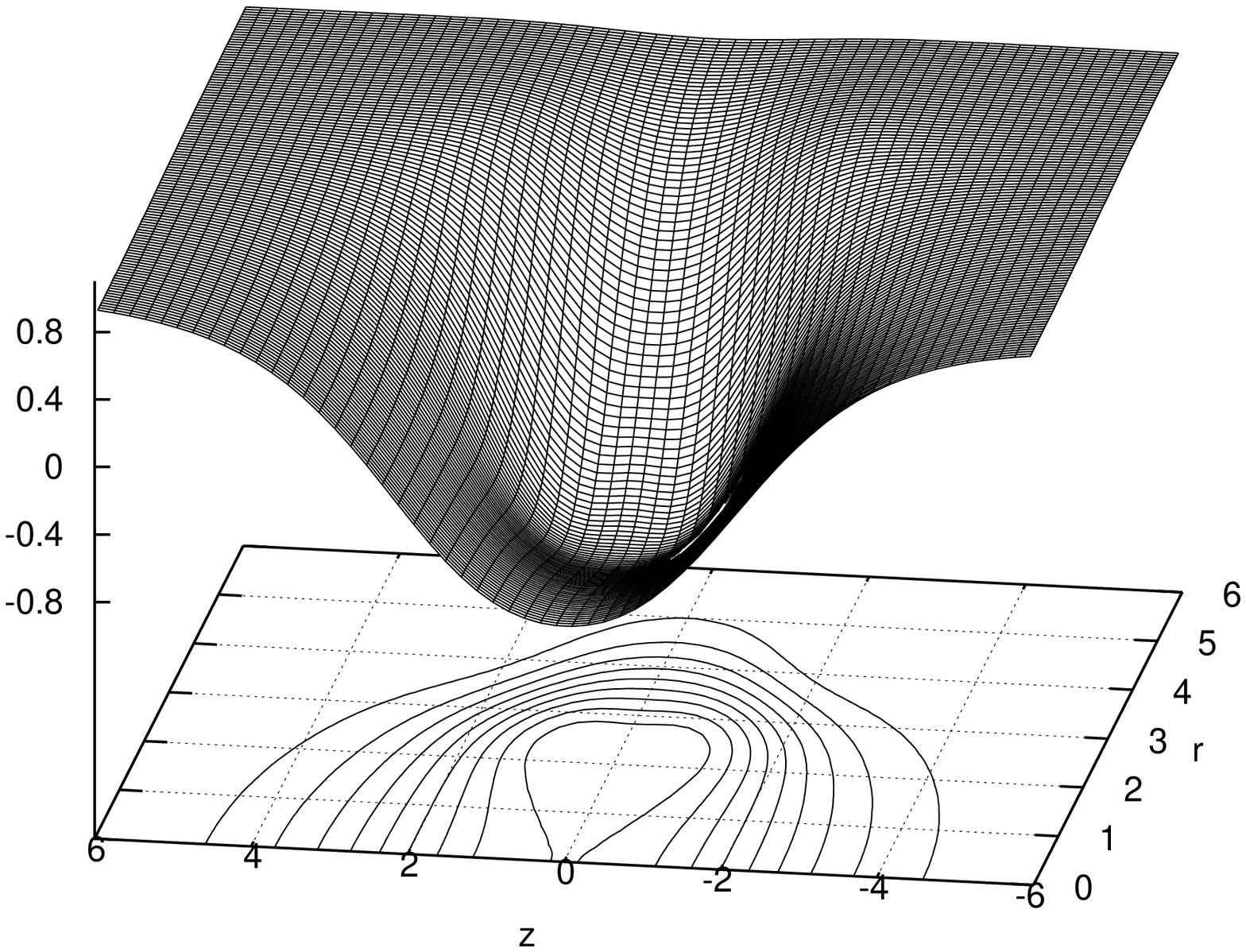}}
\end{center}
\caption{The $\sigma$ field for the $2n$ Skyrmion and charge $-2n$
antiSkyrmion solutions (right column) and its approximation from
the holonomy of charge axially symmetric $n$ calorons and charge
$-n$ anticalorons (m=2) (left column) are shown as function of the
coordinates $z$ and $\rho =\sqrt{x^2+y^2}$.}
\end{figure}

\begin{figure}
\lbfig{f-3}
\begin{center}
\subfigure[n=1]{\includegraphics[scale=0.25,angle=-90]{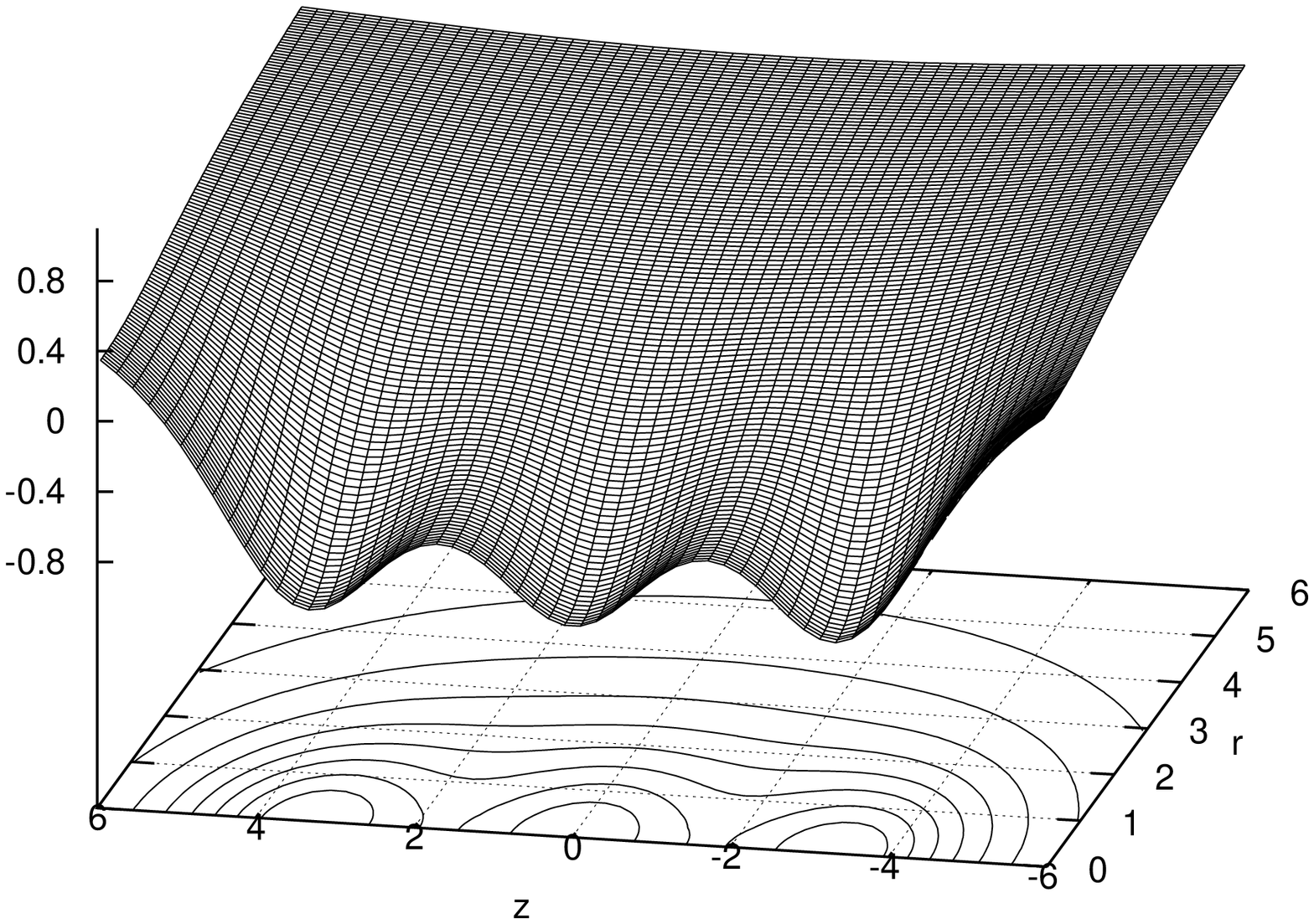}}
\subfigure[B=2n=2]{\includegraphics[scale=0.25,angle=-90]{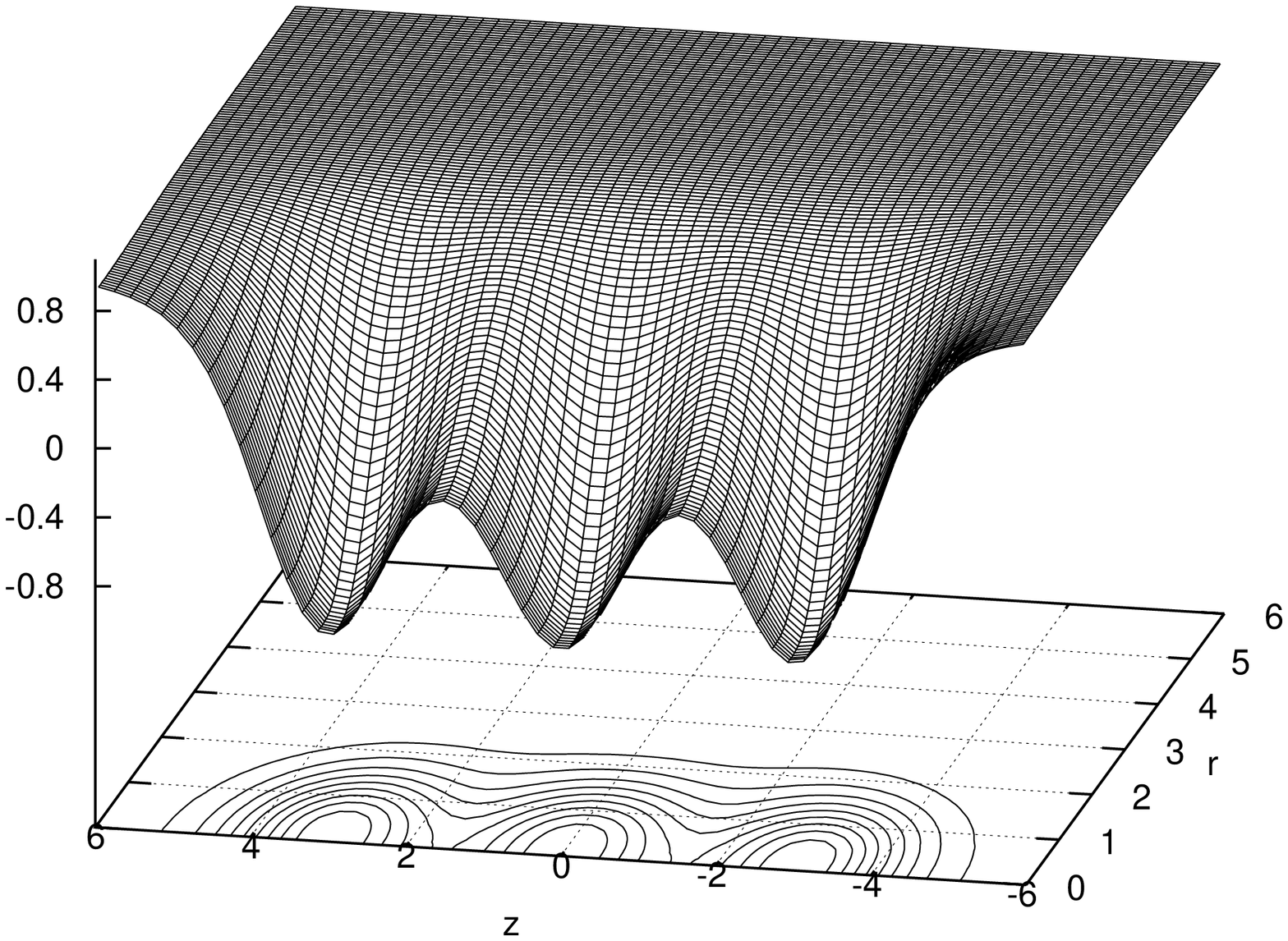}}
\subfigure[n=2]{\includegraphics[scale=0.25,angle=-90]{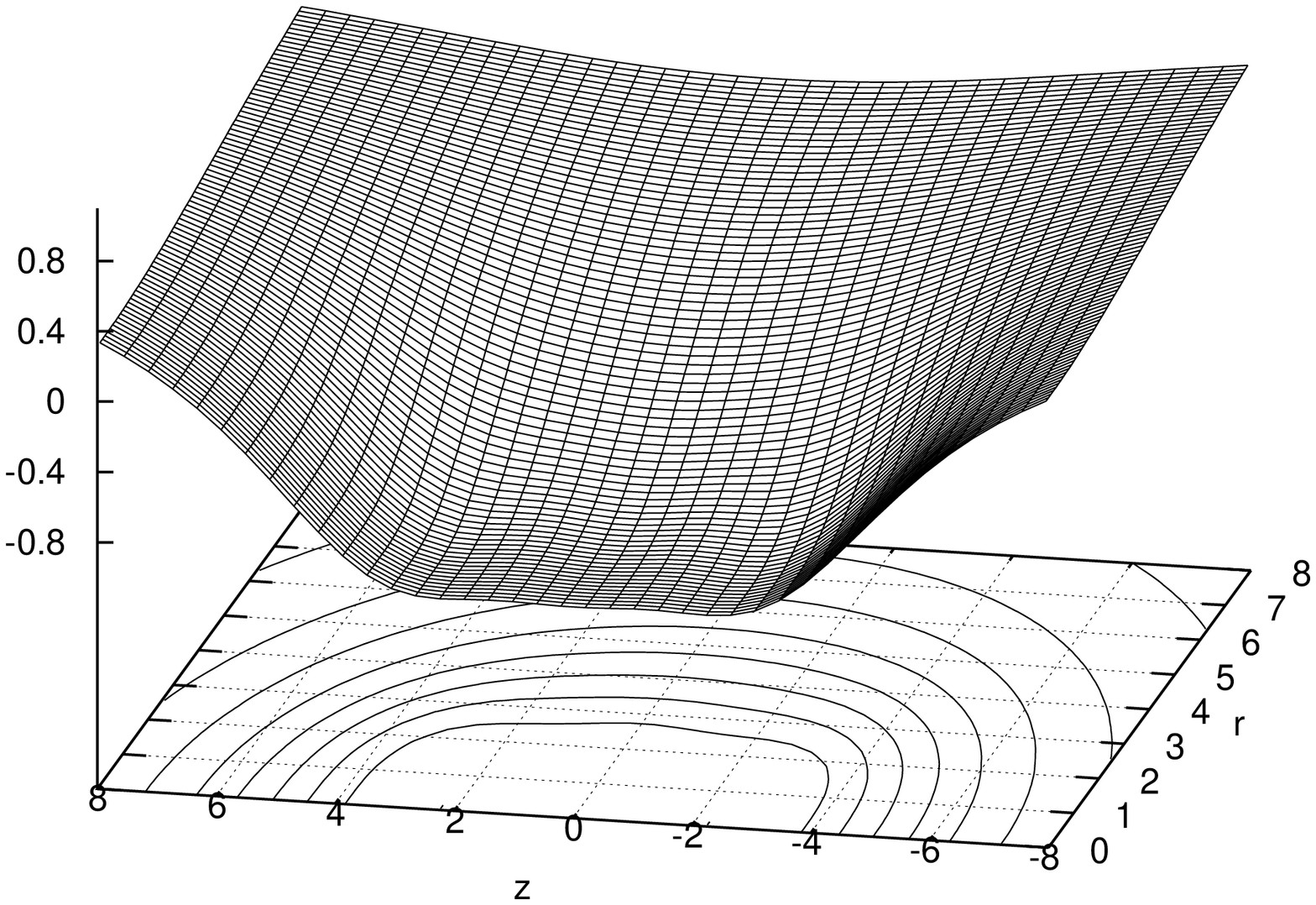}}
\subfigure[B=2n=4]{\includegraphics[scale=0.25,angle=-90]{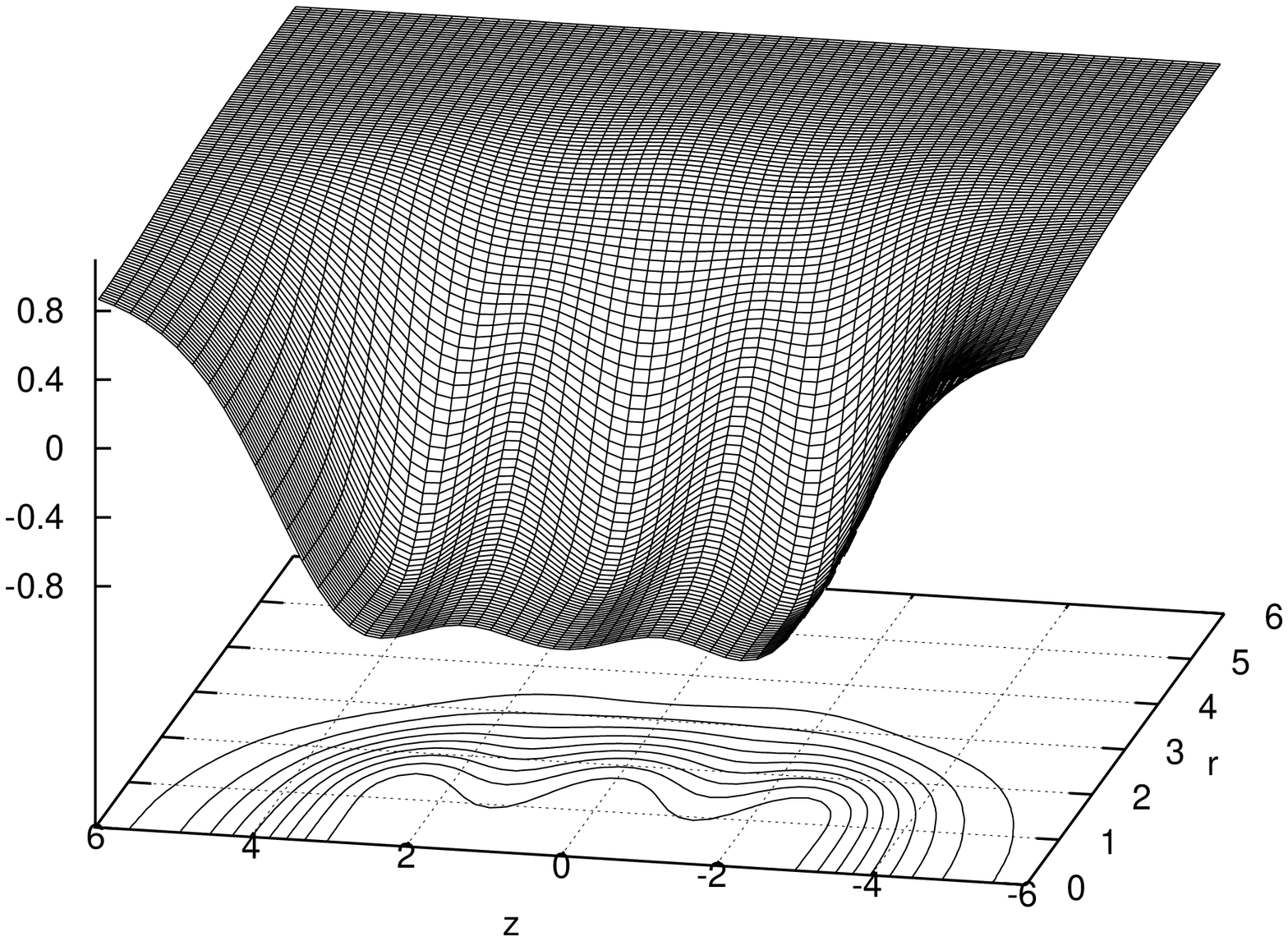}}
\subfigure[n=3]{\includegraphics[scale=0.25,angle=-90]{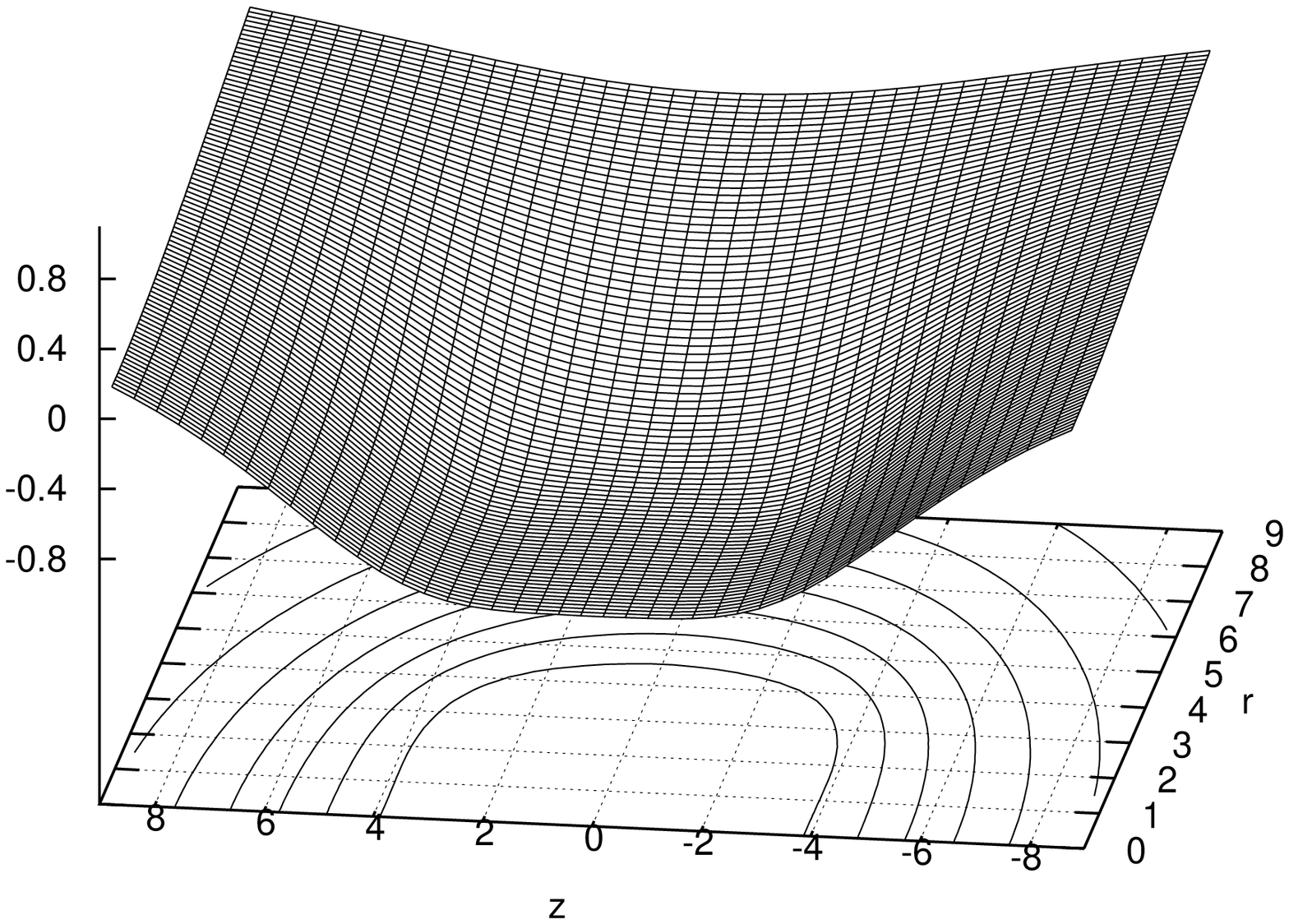}}
\subfigure[B=2n=6]{\includegraphics[scale=0.25,angle=-90]{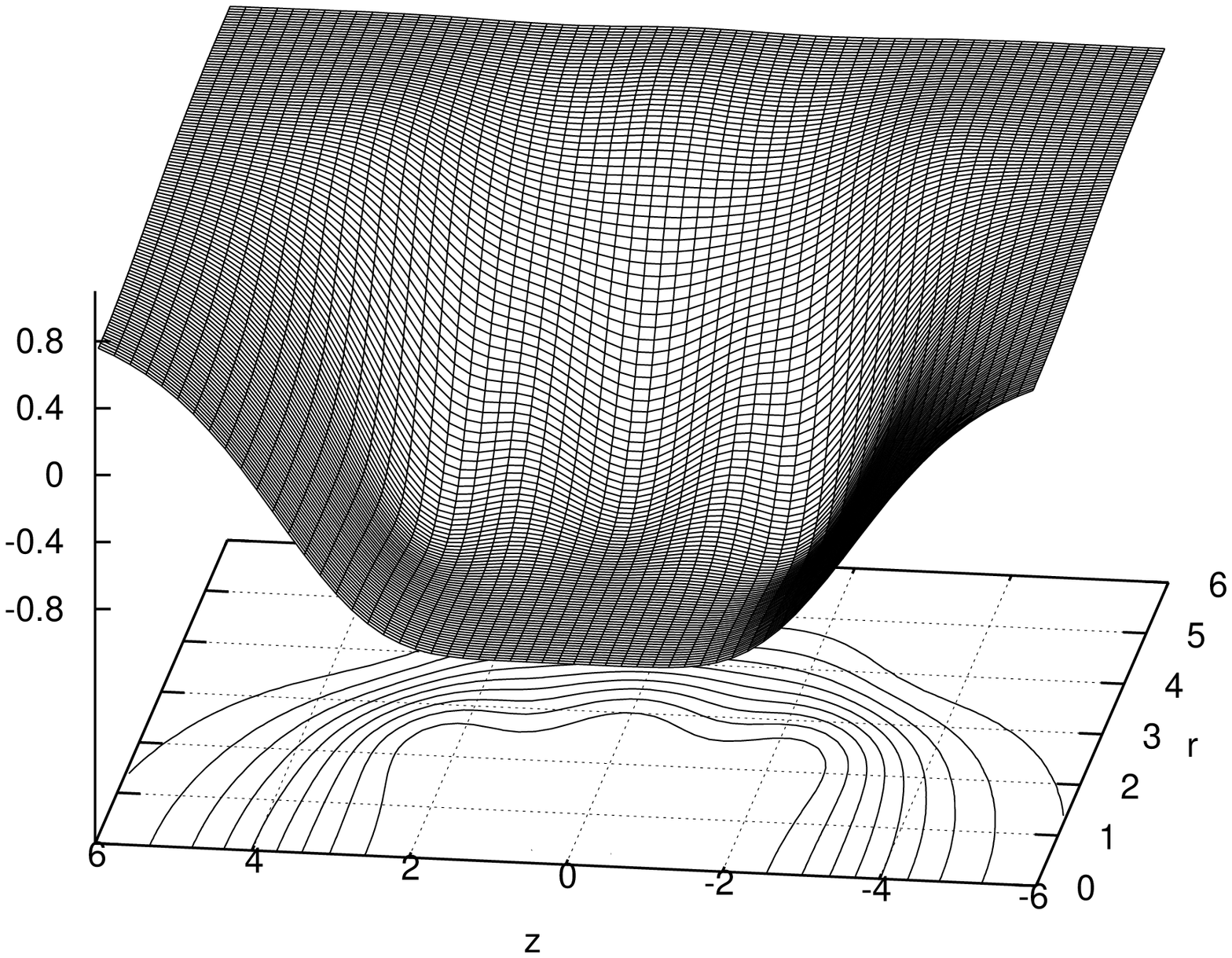}}
\end{center}
\caption{The $\sigma$ field for the $\pm 2n$ S-A-S solutions
(right column) and its approximation from the holonomy of axially
symmetric charge $n$ calorons and charge $-n$ anticalorons (m=3)
(left column) are shown as function of the coordinates $z$ and
$\rho =\sqrt{x^2+y^2}$.}
\end{figure}

It was suggested to consider a periodic version of this construction to obtain an approximation
to Skyrme crystal \cite{Manton:1995} and chains of Skyrmions \cite{Harland:2008}, then
the Skyrme field can be approximated via the
path-ordered exponential integral
 \be \label{Polakov-loop}
U({\bf r}) = {\cal P} \exp\left( \int\limits_{0}^{T}
A_0({\bf r},x_0)d x_0\right)\, , \ee
where  the period
$T$ is related with finite temperature $\Theta$ as
$T=1/k\Theta$, and $\cal P$ denotes the path ordering.

To construct Skyrmion--antiSkyrmion pairs  we shall consider deformations of the holonomy, now
the points $\tau=\pm T/4$
of each periodic interval along the Euclidean time axis should be identified with the middle point of
the double sphere which is obtained from $S^4$ compactification of ${\mathbb R}^3\times S^1$ when the latter is
twisted by $\pi$ and squeezed to a point in the equatorial hyperplane, as illustrated in Fig.~\ref{f-4}. This corresponds
to the boundary conditions we are imposing on the Yang-Mills field of the caloron-anticaloron pair \cite{KKS},
as $r\to \infty$ we have $A_\mu \, \longrightarrow \ i \partial_k U U^\dagger$ where $U = \exp\{-i \theta\tau_\varphi^{(n)}\}$, so it is element
of unity as $\theta=0$ and it is $-1$ as $\theta = \pi$.

Having constructed the  non-BPS caloron-anticaloron chains we now wish to make
use of this to compute a Skyrme field. Since we are only concerned with axially symmetric fields, the
$SU(2)$ valued Skyrme field $U({\bf r}) = \sigma \cdot {\mathbb I} + i \pi^a \cdot \tau^a$ can be parameterized as \cite{Krusch:2004uf,Shnir:2009ct}
\be
\pi^1= \phi^1 \cos(n \varphi);\quad \pi^2=\phi^1 \sin(n \varphi);\quad \pi^3=\phi^2;\quad \sigma = \phi^3
\ee
where the triplet of scalar fields $\phi^a$ on unit sphere is a function only of radial variable $r$ and polar angle $\theta$. Similar to the
axially symmetric caloron--anticaloron systems, the baryon number is $B= \frac{n}{2} \left[1-(-1)^m\right]$ where the second integer
$m$ corresponds to the number of the constituents of the configuration which can be identified with individual charge $n$ Skyrmions
and charge $-n$ antiSkyrmions \cite{Shnir:2009ct}.

For an axially-symmetric caloron-anticaloron chain configuration (\ref{ansatz}) the holonomy (\ref{Polakov-loop}) produces a Skyrme field
\begin{align}\label{psi}
\phi^1 &= \frac{1}{||A_0||}\left(K_5\sin{m\theta}+K_6\cos{m\theta}\right) \sin{\frac{T ||A_0||}{2}},\notag\\
\phi^2 &= \frac{1}{||A_0||}\left(K_5\cos{m\theta}-K_6\sin{m\theta}\right) \sin{\frac{T||A_0||}{2}},\notag\\
\phi^3 &= \cos{\frac{T||A_0||}{2}} \, ,
\end{align}
where $||A_0||=\sqrt{(K_5)^2+(K_6)^2}$. Although this construction
does not give exact solutions to the Skyrme model it does give
fields which are good approximations to the corresponding
Skyrmion-antiSkyrmion chains \cite{Krusch:2004uf,Shnir:2009ct} if
we suppose that the charge $2n$ Skyrmion and charge $-2n$
antiSkyrmion configurations corresponds to the charge $n$ caloron
and charge $-n$ anticaloron system. Indeed, Figs
\ref{f-2},\ref{f-3} demonstrate that the holonomy of the axially
symmetric charge $n$ Yang-Mills calorons and charge $-n$
anticalorons yields a good approximation to the charge $\pm 2n$
Skyrmion-antiSkyrmion (S-A) pair and to the $\pm 2n$
Skyrmion-antiSkyrmion-Skyrmion (S-A-S) configuration. Furthermore,
such a good agreement was observed for all components of the
Skyrme field (\ref{psi}). This result is not very surprising
because it is known that the the Skyrmion--antiSkyrmion chains may
exist only for value of charge $n \ge 2$
\cite{Krusch:2004uf,Shnir:2009ct}. It was also pointed out that
the transition between the chains and ring-like configurations for
the Skyrme--antiSkyrme pair it taken place as the charge $n$
increases above $n=4$ \cite{Krusch:2004uf}, in the case of
monopole-antimonopole pair similar transition occurs above $n=2$
\cite{KKS}, it indicates that the gauge interaction between the
constituents in the monopole-antimonopole pair is much stronger
than the dipole-dipole interaction in the Skyrme--antiSkyrme pair
\cite{Krusch:2004uf,Shnir:2009ct}.

Finally, let us note that the holonomy also provides a reasonable approximation to the energy of the corresponding Skyrmion--antiSkyrmion system.
\begin{figure}
\lbfig{f-1}
\begin{center}
\includegraphics[height=.50\textheight, angle =-90]{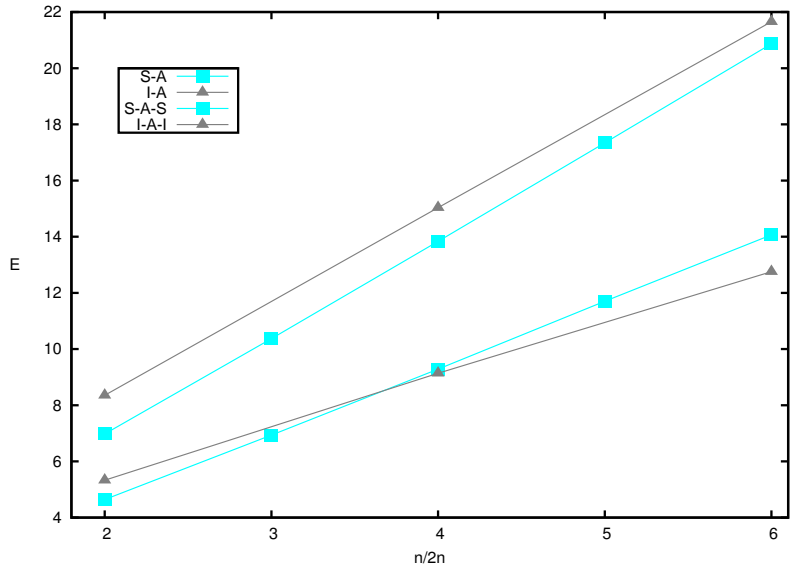}
\end{center}
\caption{
The exact energy of the Skyrmion--anti-Skyrmion chains (S-A and S-A-S)
together with the approximation energy of the corresponding caloron-anti-caloron holonomy
as functions of the topological charges of the constituents, $n$ and $2n$, respectively .}
\end{figure}
This is illustrated in Fig.~\ref{f-1}, there the curves joining the triangles
represent the holonomy approximated energy as function of the double topological charge $2n$
and the curves joining the squares represent the exact energy of the
corresponding Skyrmion--antiSkyrmion chains as function of the topological charge of the constituents $n$.

\section{Conclusions}
We have studied construction of the axially symmetric sphaleron
solutions of the Skyrme model from caloron-anticaloron holonomy
using the Atiyah-Manton approach. These configurations, both in
the Skyrme model and in the Euclidean Yang-Mills theory, are
characterized by two integers $n$ and $m$, where $\pm n$ are the
winding numbers of the constituents and the second integer $m$
defines type of the solution, the total topological charge of any
such finite action/energy configuration vanishes when $m$ is even,
and equals $n$ when $m$ is odd. This construction provides another
example of similarity which may be observed between monopoles,
instantons, calorons and Skyrmions.

We found that the caloron-anticaloron holonomy provides a good approximation to the Skyrmion-antiSkyrmion chains when
the baryon number of the constituents  is taken to be
two times more than the topological charge of the Yang-Mills calorons.
Since in that case the usual statement about the topological
equivalence of the instanton induced holonomy and the
corresponding Skyrme field cannot be applied straightforwardly,
it would be interesting to understand if there are some
topological roots of this correspondence.

Finally, let us note that the caloron-anticaloron system may exist for $n=1$, there is a difference between the
spalerons in the Yang-Mills theory at finite temperature and the  instanton–anti-instanton pair
which exists only for values $n \ge 2$ \cite{Radu:2006gg}. Since similar observation is made for the
Skyrmion-antiSkyrmion pair \cite{Krusch:2004uf}, it might be interesting to analyse
instanton-anti-instanton holonomy to generate Skyrme fields along the lines of our discussion above.

\bigskip
\noindent
{\bf\large Acknowledgements} \\
We would like to acknowledge numerous valuable discussions with Eugen Radu and Paul
Sutcliffe. YS is very grateful to E.~Norvaisas for kind hospitality at the
Institute of Theoretical Physics and Astronomy, University of Vilnius.
This work is partially supported
by BMU-MID Research Fellowships (YS).


\end{document}